\UseRawInputEncoding 
\documentclass[aps,pra,floatfix,twocolumn,superscriptaddress,longbibliography,showpacs,10pt]{revtex4-2}
\usepackage{amssymb, amsmath,mciteplus,graphicx,subfigure}
\usepackage{calc,epsfig,epstopdf,color,mciteplus,bm,mathrsfs}
\usepackage{times}
\usepackage{float}
\usepackage{lipsum}
\newcommand{\ket}[1]{| #1 \rangle} 
\newcommand{\ave}[1]{\left< #1 \right>} 
\newcommand{\ketbra}[1]{|#1\rangle\langle#1|}
\usepackage[T1]{fontenc}
\usepackage{xspace}
\usepackage{ulem}

\begin{document}
	
\title{Variational quantum simulation of the quantum critical regime}
\author{Zhi-Quan Shi}
\affiliation{Guangdong Provincial Key Laboratory of Quantum Engineering and Quantum Materials, School of Physics and Telecommunication Engineering,
South China Normal University, Guangzhou 510006, China}

\author{Xu-Dan Xie}
\affiliation{Guangdong Provincial Key Laboratory of Quantum Engineering and Quantum Materials, School of Physics and Telecommunication Engineering,
	South China Normal University, Guangzhou 510006, China}

\author{Dan-Bo Zhang}
\email{dbzhang@m.scnu.edu.cn}
\affiliation{Guangdong-Hong Kong Joint Laboratory of Quantum Matter, Frontier Research Institute for Physics,
South China Normal University, Guangzhou 510006, China} \affiliation{Guangdong Provincial Key Laboratory of Quantum Engineering and Quantum Materials, School of Physics and Telecommunication Engineering,
South China Normal University, Guangzhou 510006, China}

\date{\today}
	
\begin{abstract}
The quantum critical regime marks a zone in the phase diagram where quantum fluctuation around the critical point plays a significant role at finite temperatures. While it is of great physical interest, simulation of the quantum critical regime can be difficult on a classical computer due to its intrinsic complexity.  In this paper, we propose a variational approach, which minimizes the variational free energy, to simulate and locate the quantum critical regime on a quantum computer. The variational quantum algorithm adopts an ansatz by performing an unitary operator on a product of a single-qubit mixed state, in which the entropy can be analytically obtained from the initial state, and thus the free energy can be accessed conveniently. With numeral simulation, we show, using the one-dimensional Kitaev model as a demonstration, the quantum critical regime can be identified by accurately evaluating the temperature crossover line. Moreover, the  dependence of both the correlation length and the phase coherence time with the temperature are evaluated for the thermal states. Our work suggests a practical way as well as a first step for investigating quantum critical systems at finite temperatures on quantum devices with few qubits. 
\end{abstract}
	
\maketitle
\section{Introduction}
The quantum phase transition of quantum many-body systems marks a sharp transition between two phases and plays a central role in physics. Although occurring at zero temperature, the critical quantum fluctuation at the quantum phase transition point has far-reaching effects on the whole phase diagram, especially on a zone of quantum critical regime above the critical point that spans a range of temperatures~\cite{Sachdev}. The quantum critical regime is believed to be key for understanding a broad of physics, such as high-Tc superconductivity~\cite{lee06} and nuclear matter under finite temperature and finite density~\cite{Meyer_RMP_96,Stephanov_PRL_98}. However, the intrinsic complexity of the quantum critical regime, where both quantum fluctuation and thermal fluctuation interplay, makes a simulation of it with classical computers hard due to the sign problem~\cite{troyer2005computational}. 

In recent years, rapid advances in quantum technologies, including both quantum hardware and quantum algorithms, enable us to simulate quantum many-body systems.  It is natural to exploit current quantum processors to simulate zero-temperature quantum systems, which typically involve pure state, to investigate both static and dynamical properties of quantum systems~\cite{barends_15,bernien_17,kandala_17,zhang_17,yang_observation_2020}. Instead, simulation of finite-temperature quantum systems at equilibrium requires to prepare of a kind of mixed states called thermal states~\cite{terhal_00,poulin_09,temme_11,riera_12,wu_19,verdon_19,liu_ML_21,chowdhury2020variational,wang_PRApp_21,zhu2020generation,zhang2021continuous,xie_PRD_2022}, which can be either obtained as a subsystem of a pure state~\cite{poulin_09,wu_19,zhang2021continuous}, or as a mixture of pure states with a classical probabilistic distribution~\cite{verdon_19,liu_ML_21}. 

There are basically two approaches for thermal state preparation on a quantum computer. One is to filter out the thermal state at a temperature from a completely mixed state (infinite-temperature state) by effectively implementing an imaginary time evolution~\cite{poulin_09,mcardle_19,zhang2021continuous}. The other approach refers to variational construction of the thermal state with parameterized quantum circuit~\cite{wu_19,verdon_19,liu_ML_21,chowdhury2020variational,wang_PRApp_21,zhu2020generation}, where the optimization can be obtained by minimizing the variational free energy with a hybrid quantum-classical procedure. The variational method relies on less quantum resource and is suitable for simulating finite-temperature quantum systems on current or near-term quantum processors~\cite{Preskill_18}. However, simulation of the quantum critical regime, which demands for accurately controlling both the quantum fluctuation and thermal fluctuation and their interplay, stands out as an important goal which is less investigated. While Ref.~\cite{zhang2021continuous} has proposed to investigate the quantum critical regime with a continuous-variable assisted quantum algorithm~\cite{lau_17,zhang_20}, it should be run on a hybrid variable quantum computer which still awaits for development. In this regard, a practical way for simulating the quantum critical regime can refer to variational quantum algorithm~\cite{cerezo_variational_2021,Yuan_RMP_2020}. 

In this paper, we propose a variational approach to simulate the quantum critical regime, which is demonstrated with the one-dimensional Kitaev model under the periodic condition. The variational quantum algorithm adopts an ansatz that the variational energy can be obtained conveniently, in which the entropy is encoded in the initial state as a product of one-qubit mixed states and the parameterized unitary operator will not change the entropy. By numeral simulation, we show that such a variational quantum algorithm can prepare thermal states faithfully across the phase diagram of the Kitaev model. Remarkably, we reveal that the temperature crossover, which is important as it can locate the quantum critical regime, can be obtained accurately. We also measure
both the static and dynamic correlation functions on the optimized variational thermal states, based on which the correlation length and the phase coherence time are fitted. It is shown that both the correlation length and the phase coherence time in the critical regime for a range of intermediate temperatures are proportional to the inverse of temperature. Our work suggests that investigating quantum critical regime with few qubits can be feasible on the current quantum processors. 

\section{Quantum critical regime and variational quantum algorithm}
\label{sec:model}
In this section, we first introduce some backgrounds of the quantum critical regime with the one-dimensional Kitaev model~\cite{kitaev2001unpaired}. Then we introduce a variational quantum computing approach for simulating thermal states of the Kitaev ring~(one-dimensional Kitaev model under periodic boundary condition) as well as locating the quantum critical regime and investigating its properties . 

\subsection{Quantum critical regime of the Kitaev model}
\begin{figure}[htbp]
	\centering
	\includegraphics[width=0.8\linewidth]{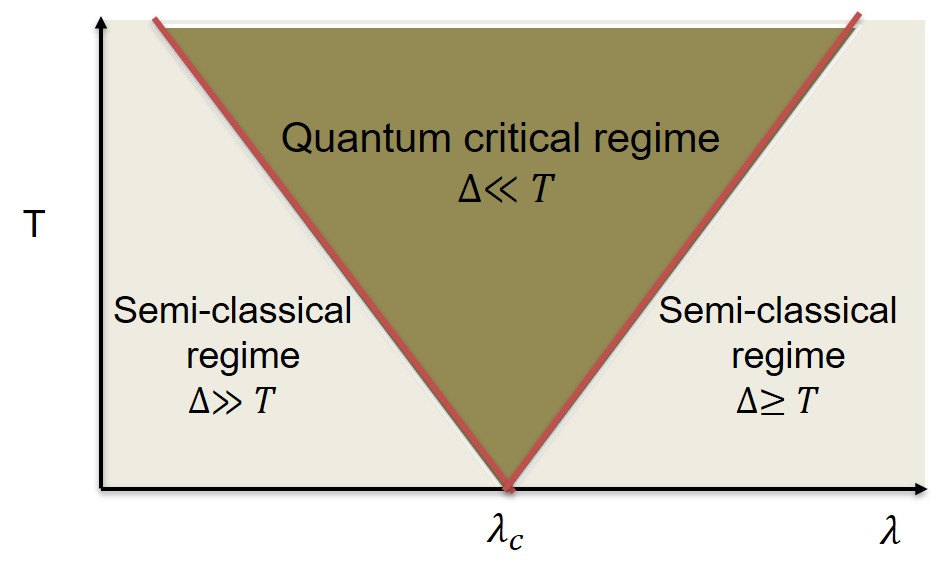}
	\caption{Illustration of the quantum critical regime. At zero temperature $T=0$, there is a phase transition point $\lambda=\lambda_c$. By comparing the energy gap $\Delta$ with the temperature $T$, the whole phase diagram can be divided into the quantum critical regime and the semi-classical regimes, which is separated by the temperature crossover line~(red lines).  }\label{Fig:illustration}
\end{figure}

Quantum phase transition is defined at zero temperature where the phase of state dramatically changes when tuning a parameter of the system across a point. The critical point associated with the quantum phase transition, although occurs at zero temperature, has far-reaching effects for the phases of state at finite temperature~\cite{Sachdev}.
By comparing two important energies scales of the system, namely the gap $\Delta$ and the temperature $T$, the phase diagram can be divided into regimes $\Delta\gg T$ and $\Delta\ll T$, as shown in Fig.~\ref{Fig:illustration}. The regime of $\Delta \gg T$ represents low-$T$ regime where dynamics and transport can be described in a semi-classical way. The regime of $\Delta \ll T$, where both quantum fluctuation and classical fluctuation interplay, marks a quantum critical regime, which is separated from the semi-classical regimes with the temperature crossover lines. The quantum critical regime owns universal properties, such as the correlation length and the phase coherence time have scaling behaviors with the temperature~\cite{Sachdev}. 

The quantum critical regime plays a key role in understanding a broad of physics such as high-Tc superconductivity and nuclear matter. The difficulty of studying quantum critical systems lies in the intrinsic complexity associated with thermals states that describe phases of state in the quantum critical regime~\cite{troyer2005computational}.   
Quantum simulation can directly prepare those thermal states and measure the physical properties. In this regard, it provides a bottom-up approach to investigating the quantum critical regime. However, as investigated in our previous work~\cite{zhang2021continuous}, a basic goal of simulating the quantum critical regime by locating the temperature crossover line can be tricky as it is model-dependence. For instance, it typically requires more than a system size of more than $100$ sites for the quantum Ising model, and on the other hand, it demands only a few sites for the Kitaev ring~\cite{zhang2021continuous}. Thus, we chose the Kitaev ring as a model Hamiltonian for simulating the quantum critical regime. 

The Hamiltonian of the Kitaev ring model reads
\begin{eqnarray}
\label{ham_ks}
H_{K}={-J} \sum_{i=1}^{N}[{c }_{i}^{\dagger }{c}_{i+1}+{c }_{i}^{\dagger }{c}_{i+1}^{\dagger }+h.c.]{-u} \sum_{i=1}^{N}{c }_{i}^{\dagger }{c}_{i+1},
\end{eqnarray}
where fermion operators ${c}_{N+1}={c}_{1}$ as imposed by the periodic condition. The Kitaev model has a quantum phase transition at $\frac{u}{2J}=1$. To simulate the Kitaev ring model, the fermion operator should be mapped to qubit operator by the Jordan-Wigner transformation, ${c}_{i}=\prod\limits_{j=1}^{i-1}{\sigma}_{j}^{z}{\sigma}_{i}^{-}$,${c}_{i}^{\dagger}={\sigma}_{i}^{\dagger}\prod \limits_{j=1}^{i-1}{\sigma}_{j}^{z}$, ${c}_{i}^{\dagger}{c}_{i}=\frac{1}{2}({\sigma}_{i}^{z}-1)$.
Now the spin Hamiltonian reads as~(omitting a constant), 
\begin{eqnarray}
\label{ham_ks}
H=-J\sum_{i=1}^{N-1}{\sigma}_{i}^{x}{\sigma}_{i+1}^{x}-J{\sigma}_{1}^{y}P{\sigma}_{N}^{y}{-\lambda} \sum_{i=1}^{N}{\sigma}_{i}^{z},
\end{eqnarray}
where  $\lambda=\frac{u}{2}$, $P=\prod\limits_{i=2}^{N-1}{\sigma}_{i}^{z}$ is a string operator. Hereafter we set $J=1$. The model in Eq.~\eqref{ham_ks} is in fact close to the transversal field Ising model~(TFIM),  except that the boundary term ${\sigma}_{1}^{y}P{\sigma}_{N}^{y}$is different, e.g., the term should be ${\sigma}_{1}^{x}{\sigma}_{N}^{x}$ for TFIM. 

We now turn to discuss the Kitaev ring at finite temperature. The quantum system of the Kitaev ring at equilibrium under an inverse temperature $\beta=1/T$ can be described by a thermal state~(also known as Gibbs state)~\cite{kardar_2007}, $\rho(\beta)=e^{-\beta H}/Z(\beta)$, where $Z(\beta)=\text{Tr}e^{-\beta H}$ is the partition function. The free energy is related to the partition function as $F(\beta)=-\beta^{-1}\ln{Z(\beta)}$. The free energy at parameter $\lambda$ is denoted as $F(\beta,\lambda)$.  

The thermodynamic properties can be derived from the free energy or by measuring observable the thermal states. One conventional approach to locate the temperature cross lines is to calculate the magnetic susceptibility and identify the temperature cross point $T^*(\lambda)$ for a given $\lambda$ as a temperature where the susceptibility is maximum~\cite{Sachdev,zhang2021continuous}. From statistical physics, it is known that the magnetic susceptibility can be expressed as
\begin{equation}\label{eq:chi_def}
	\chi(\beta,\lambda)=\frac{\partial^2F(\beta,\lambda)}{\partial^2\lambda}.
\end{equation} Thus, the temperature cross point for a given $\lambda$ is, 
\begin{equation}\label{eq:T_cross_def} T^*=\arg\max_T\chi(1/T,\lambda). 
\end{equation} 
It should be pointed out that the quantum critical regime would not include the zone $T>J$, which is dominated by lattice cutoff and the properties are not universal~\cite{Sachdev}.

Another important aspect of the quantum critical regime is 
the scaling behavior of the correlation length $\xi$ and the phase coherence time $\tau$~\cite{Sachdev}. For the spin chain of $H$, it is known that both $\xi$ and $\tau$ are proportional to the inverse temperature $\beta$ in the quantum critical regime. The correlation length and the phase coherence time should be evaluated from the static correlation function and the dynamical correlation function, respectively. The static correlation function is defined as
\begin{equation}\label{eq:R_def}
	R(n)=\sum_{i=1}^{N}\text{Tr}[\rho(\beta)\sigma^x_i\sigma^x_{i+n}].
\end{equation}
At nonzero temperature, $R(n)$
should be exponentially decreasing with the spatial separation $n$ . The correlation length is defined as a characterization length by $R(n)\propto e^{-n/\xi}$. The dynamical correlation function can be chosen as
\begin{equation}\label{eq:C_def}
C(t)=\sum_{i=1}^{N}|\text{Tr}[\rho(\beta)\sigma^x_i(t)\sigma^x_{i}]|,
\end{equation}
where $\sigma^z_i(t)=e^{iHt}\sigma^z_ie^{-iHt}$. Note that in the summation each dynamical correlation function takes absolute value as it is a complex number. Similarly,  $C(t)$ at finite temperature is exponential decreasing with $t$ and the phase coherence time is defined through $C(t)\propto e^{-t/\tau}$.      

\subsection{Variational quantum algorithm}
We now propose a variational quantum computing approach for simulating the quantum critical regime for the Kitaev ring. This includes two goals: locating the quantum critical regime and investigating the scaling behavior. Reaching those goals relies on preparing thermal states accurately on a quantum computer, based on which physical quantities can be evaluated reliably. 

\subsubsection{Complexity of thermal states}\label{subsub:complexity}
Let us first illustrate the complexity of preparing thermal states. It is inspiring to decompose the thermal state as, 
\begin{equation}
	\rho(\beta)=\sum_{i=1}^{2^N}p_i(\beta)\ketbra{\psi_i},~~ p_i(\beta)=\frac{e^{-\beta E_i}}{Z(\beta)},
\end{equation}   
where $H\ket{\psi_i}=E_i\ket{\psi}$. The thermal state thus is a mixture of eigenstates $\{\ket{\psi_i}\}$ with classical probabilities $\{p_i\}$. At low temperatures, only the ground state has a large weighting and the task is almost reduced for preparing the ground state. However, for higher temperatures, such as $T \sim \Delta$, the low-lying eigenstates will have large probabilities and it requires to prepare accurately those low-lying eigenstates $\ket{\psi_i}$ with the corresponding weights $p_i$ at the same time. Such a task is harder than preparing the ground state. However, for very high temperatures the task of preparing the thermal state becomes easy. To see this, we consider the infinite-temperature limit $\beta=0$. The thermal state is a completely mixed state $\rho(\beta=0)=I/2^N$. As $U\rho(\beta=0)U^\dagger=\rho(\beta=0)$, where $U$ is a unitary transformation, $\rho(\beta=0)$ is in fact an equal-mixing of an arbitrary set of complete basis $\{U\ket{\psi}\}$. Another aspect to reveal the simplicity of $\rho(\beta=0)$ is to that it equals to a product of single-qubit mixed state $\otimes_{i=1}^{N} I/2$.  In other words, there is no correlation at $\beta=0$ and the temperature is local~\cite{kliesch2014locality}. In fact, it can be proved theoretically that the correlation length scales as $\xi(\beta)=\beta^{\frac{2}{3}}$ for local Hamiltonian at high temperatures~\cite{Kuwahara_PRX_2021}. Since it requires a deeper quantum circuit for preparing states with longer correlation length~\cite{ho2019efficient,Kuwahara_PRX_2021}, such a scaling indicates that the complexity of preparing a thermal state reduces when increasing the temperature at the high-temperature regime. 

The above discussion suggests that preparing thermal states at intermediate temperatures will be harder than low-temperature and high-temperature ones. This is just the case for locating the temperature crossover lines which connect the regime of $\Delta\ll T$ and $\Delta\gg T$. In this regard, it can be challenging to simulate the quantum critical regime for quantum computing. 

\subsubsection{Variational preparing thermal states}
The variational quantum computing approach can meet this challenge by taking the advantage of using a specific ansatz that may require short-depth quantum circuit and thus can be suitable on near-term quantum processors. 
The variational principle for the quantum system at temperature $T$ is that the free energy should be minimized for the thermal state~\cite{kardar_2007}. One can prepare a variational thermal state $\rho(\boldsymbol{\omega},\beta)$ with a parameter set $\boldsymbol{\omega}$ on a quantum computer. The  $\rho(\boldsymbol{\omega};\beta)$ can be generated either as a subsystem of a pure state or as a mixture of pure states. The parameter set $\omega$ should be optimized by minimizing the variational free energy expressed as,     
\begin{eqnarray} \label{eq:F_E_S}
	F(\boldsymbol{\omega};\beta)=E(\boldsymbol{\omega})-TS(\boldsymbol{\omega}),
\end{eqnarray}
where $E(\boldsymbol{\omega})=\text{Tr}[\rho(\boldsymbol{\omega};\beta)H]$ is the average energy and  $S(\boldsymbol{\omega})=-\text{Tr}[\rho(\boldsymbol{\omega};\beta)\log\rho(\boldsymbol{\omega};\beta)]$ is the von Neumann entropy. The energy $E(\boldsymbol{\omega})$ can be evaluated by decomposing the Hamiltonian as a linear combination of local observable and 
measuring each separately. However, estimating the von Neumann entropy  is a difficulty task in general since it does not associate with a Hermitian observable. Indeed some quantum protocols can measure the Reyi entropy which is a function of $\rho^n$~($n$ is a positive integer)~\cite{klich2006measuring,islam2015measuring,brydges2019probing}. However, measuring the von Neumann entropy is more challenging as it involves $\log\rho$ and there is still lack of efficient protocol for generic quantum states, except for some proposals with approximation valid under specific conditions ~\cite{audenaert2007sharp, acharya2019measuring,wang_PRApp_21}. 

\begin{figure}[htbp]
	\centering
	\includegraphics[width=1\linewidth]{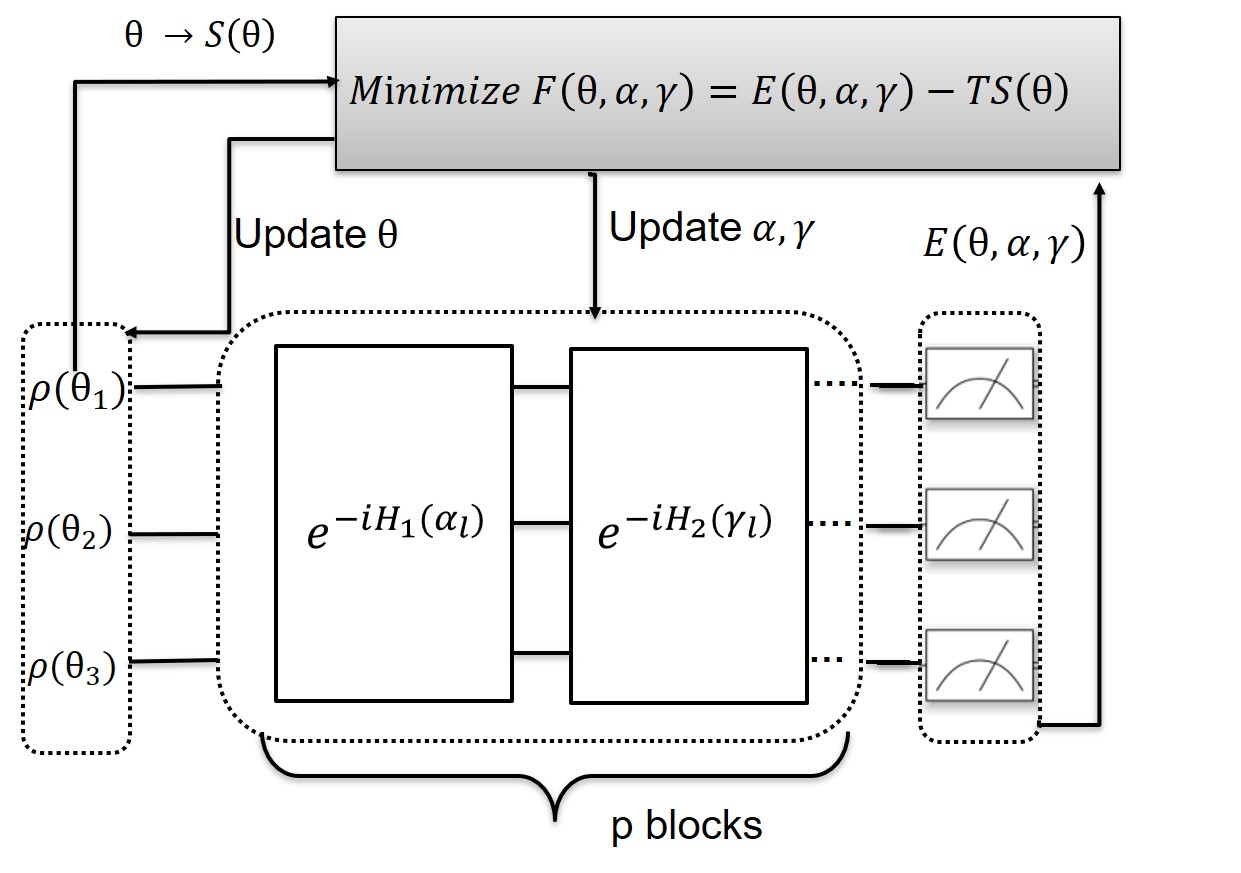}
	\caption{Illustration of the variational quantum algorithm. The variational Gibbs state is prepared by performing a parameterized unitary operator on an initial state $\otimes_i\rho_i(\theta_i)$. The parameterized circuit consists of $p$ blocks. By measuring the energy $E$ on a quantum computer by calculating the entropy $S$ from $\theta$, the variational free energy $F$ can be obtained. With a hybrid quantum-classical optimization, parameters ($\theta, \alpha,\gamma)$) are iteratively updated to minimize the variational free energy. }\label{Fig:illustration_circuit}
\end{figure}

One solution is to use some specific ansatz so that the von Neumann entropy can be calculated directly without measurements. This is possible by using an ansatz where the classical probability and the eigenstates are parameterized separately~\cite{martyn2019product,liu2021solving,xie_PRD_2022}, e.g., the variational thermal state can take a formula~(with a parameter set $\boldsymbol{\omega}=(\theta,\phi)$), 
\begin{equation}\label{eq:PSA_ansatz}
\rho(\boldsymbol{\omega};\beta)=\sum_{i=1}^{2^N}p_i(\theta)U(\phi)\ketbra{i}U^\dagger(\phi)=U(\phi)\rho_0(\theta)U^\dagger(\phi), 
\end{equation}
where $\rho_0(\theta)=\sum_{i=1}^{2^N}p_i(\theta)\ketbra{i}$. The unitary evolution $U(\phi)$ will not change the entropy of the initial state $\rho_0(\theta)$. Thus the entropy can be obtained from the classical probability $\{p_i(\theta)\}$,
\begin{equation}
	S(\theta)=-\sum_{i=1}^{2^N}p_i(\theta)\log p_i(\theta). 
\end{equation}

Following Refs.~\cite{martyn2019product,xie_PRD_2022}, we chose the initial state as a product state,
\begin{eqnarray}
	\label{eq:initial}
	\rho_0(\theta)&=&\otimes_{i=1}^{N}\rho_i(\theta_i),\nonumber \\
\text{where}~	\rho_i(\theta_i)&=&\sin^2\theta_i\ketbra{0_i}+\cos^2\theta_i\ketbra{1_i}.
\end{eqnarray} 
Such a choice of initial state together with Eq.~\eqref{eq:PSA_ansatz} is known as the product-spectrum ansatz~\cite{martyn2019product}.
Calculation of the entropy can be simplified as it is a summation of entropy for each qubit,
\begin{equation}
	S(\theta)=\sum_{i=1}^{N}[-\sin^2\theta_i\log\sin^2\theta_i-\cos^2\theta_i\log\cos^2\theta_i].
\end{equation}
While using only $N$ parameters to characterize the classical probability, it is shown that the product-spectrum ansatz can faithfully represent thermal states of a broad of physical systems~\cite{martyn2019product,xie_PRD_2022}.  
It is noted that the single-qubit mixed state $\rho_i(\theta_i)$ can be obtained by tracing one qubit of a two-qubit pure state 
$\sin\theta_i\ket{00}+\cos\theta_i\ket{11}$. This takes an overall $2N$ qubits to prepare the thermal state. One may also take $\rho_0(\theta)$ as a mixture of computational basis $\ket{s_1s_2...s_N}$, where $s_i=0,1$. If $N$ is small, $\rho_0(\theta)$ can be generated by preparing an initial state $\ket{s_1s_2...s_N}$ with a probability $\prod_{i=1}^{N}f_{s_i}(\theta_i)$, where $f_0=\sin$ and $f_1=\cos$. The probabilistic way uses only $N$ qubits but involves an ensemble of quantum circuits whose number increases exponentially with $N$. 

The unitary operator $U(\phi)$ can be constructed with different types of parameterized quantum circuits. Here we use the following structure involving $p$-blocks of alternatively Hamiltonian evolution~(with a parameter set $\phi=(\alpha,\eta)$)~\cite{xie_PRD_2022,Chen_CPL_2013}, 
\begin{eqnarray}
	\label{Eq:unitary} 
	U(\phi)\equiv U(\alpha,\eta)=\prod_{l=1}^{p} e^{-iH_{2}(\eta_{l})}e^{-iH_1({\alpha_{l}})}
\end{eqnarray}
where $H_{1}(\alpha_{l})=\sum_{i=1}^{N}\alpha_{l,i}\sigma^z_{i}$ and $H_{2}(\eta_{l})=\sum_{i=1}^{N-1}\eta_{l,i}\sigma_{i}^{x}{\sigma}_{i+1}^{x}+\eta_{l,N}{\sigma}_{1}^{y}P{\sigma}_{N}^{y}$. As terms in $H_{1}(\alpha_{l})$ commute to each other, $e^{-iH_1({\alpha_{l}})}$ can be directly decomposed as a series of quantum gates. The same situation applies for $e^{-iH_{2}(\eta_{l})}$. The choice of $U(\phi)$ is physical motivated as $H_{1}(\alpha_{l})$ and $H_{2}(\eta_{l})$ inherit from the Hamiltonian $H$. While similar to the Hamiltonian ansatz~(HVA)~\cite{wiersema-PRXQuantum2020}~(also known as Quantum Alternating Operator Ansatz \cite{Hadfield2017FromTQ}), it allocates each term with a variational parameter, rather than allocating the same parameter to all terms. We may call the unitary in Eq.~\eqref{Eq:unitary} as multi-angle HVA~\cite{Chen_CPL_2013}. The original HVA is proposed to prepare the ground state. A promotion of HVA to the multi-angle HVA is necessary as preparing thermal state is more difficulty and demands more representation power of the ansatz. 

With the initial state $\rho_0(\theta)$ in Eq.~\eqref{eq:initial} and the unitary operator $U(\alpha,\eta)$ in Eq.~\eqref{Eq:unitary}, the variational state can be written as $\rho(\theta,\alpha,\eta;\beta)$. To optimize the parameter $(\theta,\alpha,\eta)$, one can minimize the variational free energy $F(\theta,\alpha,\eta;\beta)$ using a hybrid quantum-classical procedure. The parameter for the initial state $\rho(\theta)$ and the parameter $(\alpha,\eta)$ in the quantum circuit are updated respectively after evaluating the free energy in each iteration.  An illustration of the variational quantum algorithm for preparing the thermal state as well as the hybrid quantum-classical optimization is given in Fig.~\ref{Fig:illustration_circuit}.

\subsubsection{Evaluation of physical quantities}

With the optimized variational free energy, we now can evaluate the susceptibility in Eq.~\eqref{eq:chi_def} by a second-order difference scheme, 
\begin{equation}\label{eq:diff_chi}
	\chi(\beta,\lambda)\approx \frac{F(\beta,\lambda+\delta\lambda)+F(\beta,\lambda-\delta\lambda)-2F(\beta,\lambda)}{(\delta\lambda)^2},
\end{equation} 
where $\delta\lambda$ is a small number and variational parameters in $F$ have not been written explicitly. For each $\lambda$, a series of $\chi(\beta,\lambda)$ are calculated, and the temperature crossover point is identified as a temperature that $\chi(\beta,\lambda)$ is maximum as in Eq.~\eqref{eq:T_cross_def}. By going through all $\lambda$, the temperature crossover lines $T(\lambda)$ can be identified. 

\begin{figure}[htbp]
	\centering
	\includegraphics[width=0.9\linewidth]{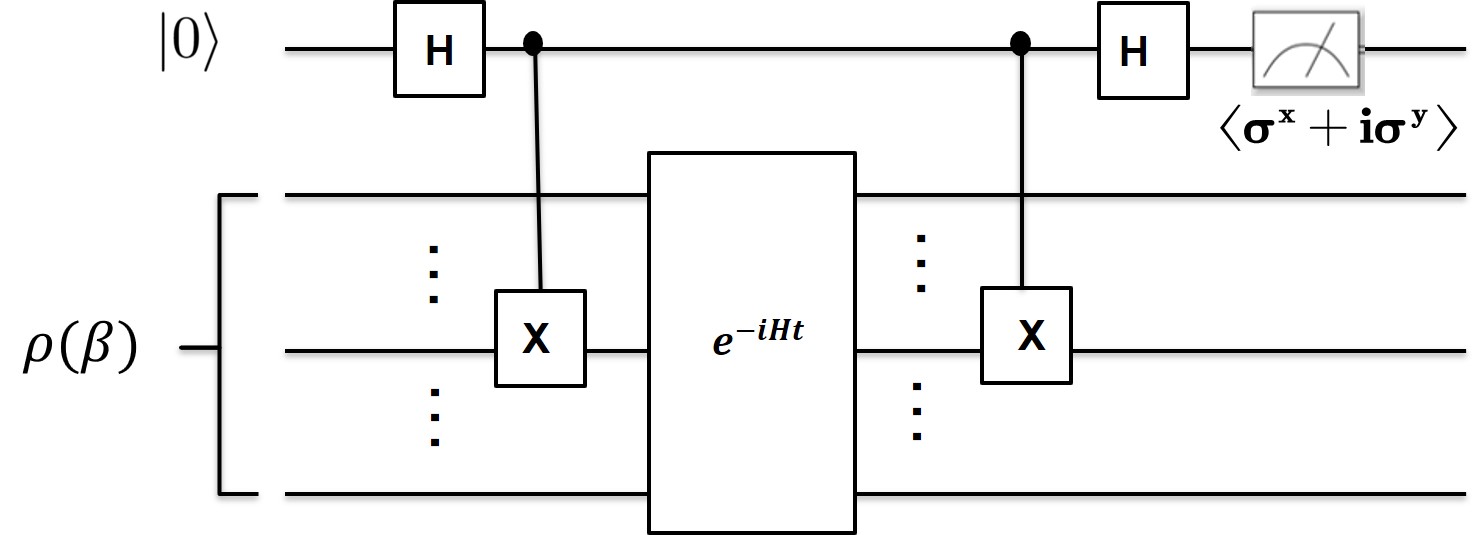}
	\caption{Quantum circuit for computing the dynamical correlation function $\text{Tr}[\rho(\beta)\sigma^x_i(t)\sigma^x_{i}]$. The first qubit initialed as $\ket{0}$ is an ancillary qubit. }\label{Fig:circuit_dynamical}
\end{figure}

With the optimized variational thermal state, the spatial and dynamical correlation function can be measured as in Eq.~\eqref{eq:R_def} and Eq.~\eqref{eq:C_def}, respectively. Evaluating the spacial correlation function $R(n)$ relies on a joint-measurement of two qubits. By measuring $R(n)$ at different spacing $n$, the correlation length $\xi$ is obtained by fitting $R(n)\propto e^{-n/\xi}$. Measuring the dynamical correlation function $C(t)$ is also a standard technique~\cite{Pedernales_PRL_14,Li_PRD_2022}. The corresponding quantum circuit is given in Fig.~\ref{Fig:circuit_dynamical}.
The value of $C(t)$ is recorded by measuring $\ave{\sigma^x+i\sigma^y}$ on the ancillary qubit. Similarly, by measuring $C(t)$ at different time $t$, the phase coherent time $\tau$ can be obtained by fitting $C(t)\propto e^{-t/\tau}$. 

It should be emphasized that studying the scaling behavior should refer to the thermodynamic limit where the system size is infinite. As quantum simulation can only be performed on finite-size systems, one may conduct finite-size scaling~\cite{Fisher_PRL_72}. In this work, however, we only do some primary investigations on the scaling behavior without sophisticated finite-size scaling analysis due to limited simulation capacity.   

\section{Results}\label{sec:results}
In this section, we represent simulation results. The simulation is performed with the open-source package $\it{projectQ}$~\cite{projectQ} on classical computers. We use BFGS for the optimization, which is a gradient-based method. We also adopt a strategy to boost the optimization by utilizing a continuous relation between optimized variational parameters with the temperature~\cite{zhang_PRA_2020,Yuan_PRA_2021}. Concretely, we first get optimized variational parameters for high-temperature thermal states as they are comparatively easy to solve. The optimized parameters then are set as initial parameters of the thermal state of lower-temperature system. The procedure is repeated when the temperature is reduced to zero temperature. With this strategy, the thermal states of the whole phase diagram can be obtained more efficiently. 

\begin{figure}[htbp]
	\centering
	\includegraphics[width=1\linewidth]{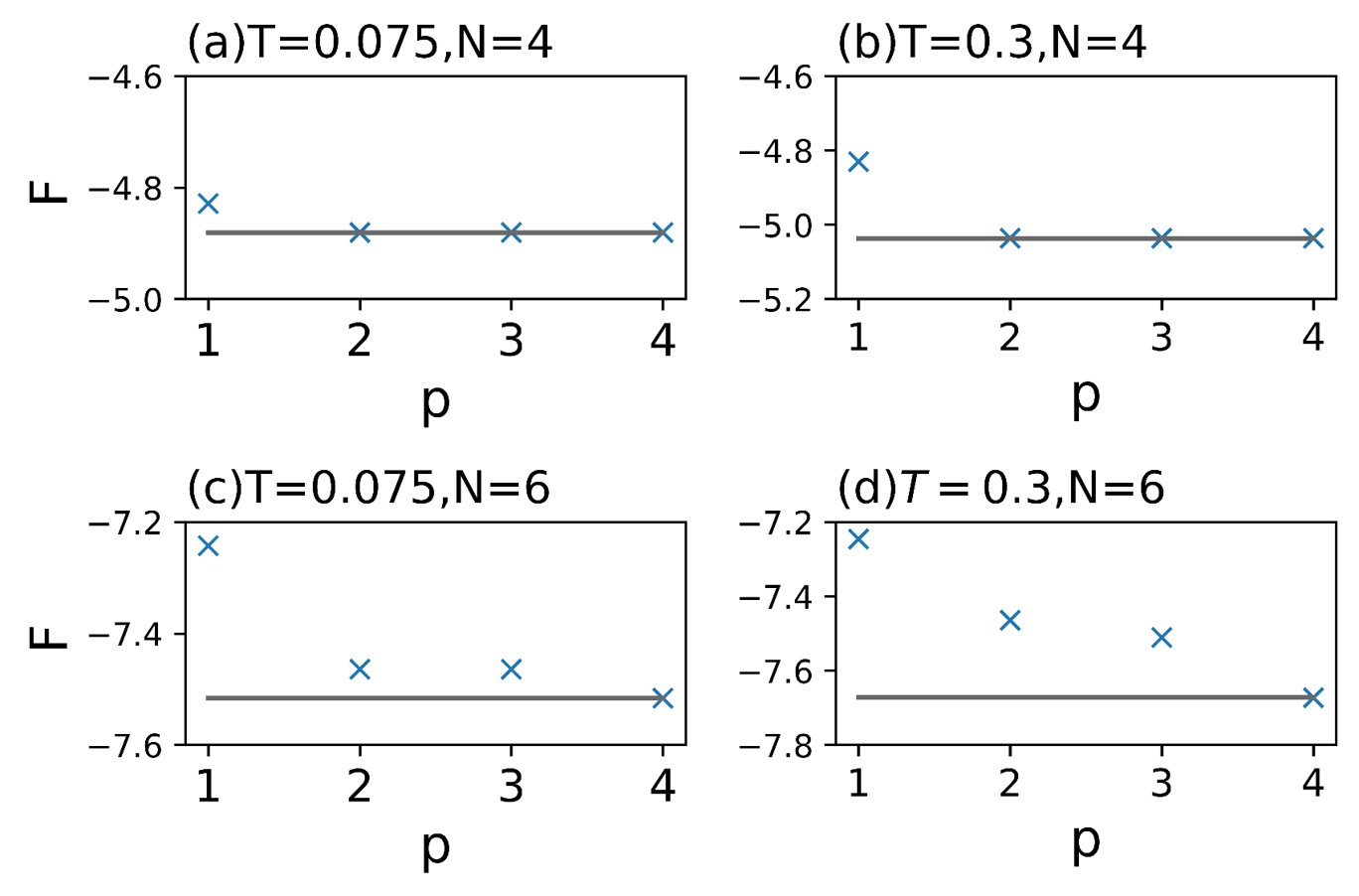}
	\caption{Numeral simulation results~(scatters) of free energy are compared with exact ones~(lines) by increasing the number of blocks $p$ in the quantum circuit.}\label{Fig:F_with_p}
\end{figure}

\begin{figure}[htbp]
	\centering
	\includegraphics[width=1\linewidth]{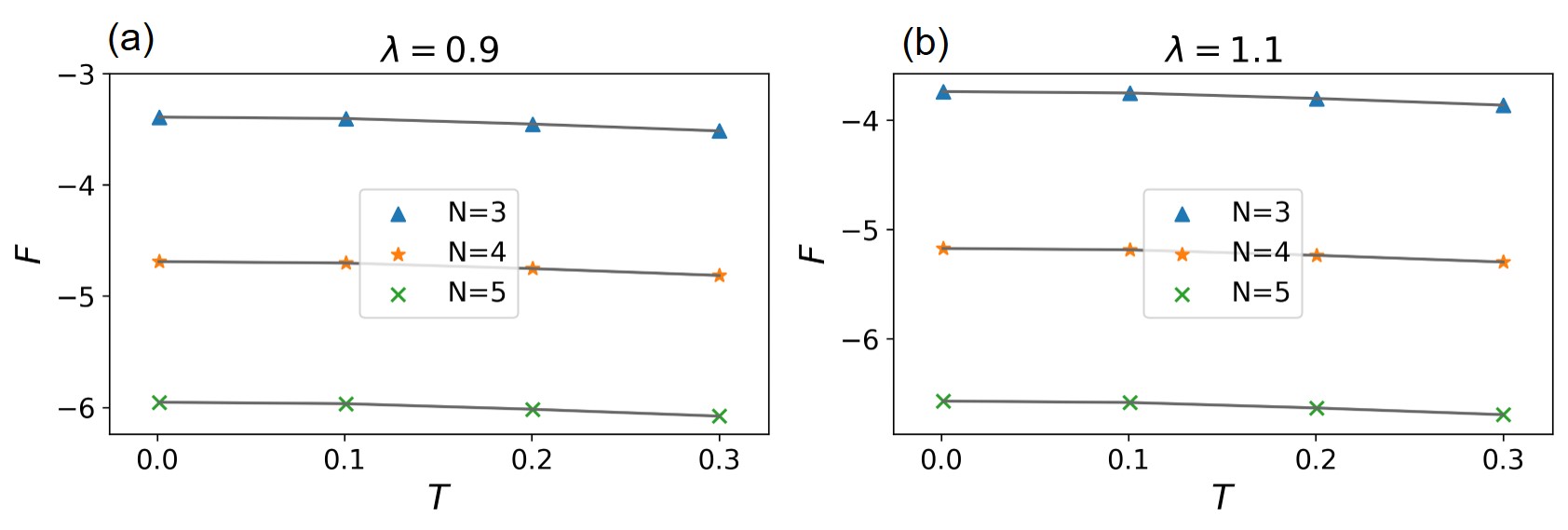}
	\caption{Numeral simulation results~(dots) of free energy are compared with exact ones~(lines) with increasing temperature. The number of blocks is $p=5$.  }\label{Fig:F_with_T}
\end{figure}

We test the variational algorithm for preparing the thermal states of the Kitaev ring by comparing the optimized free energy with the exact ones. Firstly, the accuracy for solving thermal states at different system sizes is investigated with increasing circuit depth characterized by the number of blocks $p$ in the unitary operator Eq.~\eqref{Eq:unitary}. As seen Fig.~\ref{Fig:F_with_p}, with increasing $p$, the optimized free energy will converge nearly to the exact one. Moreover, a larger-size system will require a larger $p$ for obtaining accurate free energy. The demanding of more quantum resources for simulating larger quantum systems at finite temperatures is expected but the exact scaling of quantum resources with the system size is still an open question. For the transversal field Ising model, it has been argued and numerically verified that the critical point requires a depth $O(N)$ to prepare the ground state with the HVA~\cite{ho2019efficient}. For the multi-angle HVA, the requirement of $p$ with the system size $N$ is still awaited investigation. Secondly, the accuracy of free energy at different temperatures is shown in Fig.~\ref{Fig:F_with_T}. We choose $p=5$ for all cases.  For different sizes~($N=3,4,5$) and different $h$~($h=0.9,1.1$), it is shown that the accuracy is good for the whole temperature range. 


\begin{figure}[htbp]
	\centering
	\includegraphics[width=1\linewidth]{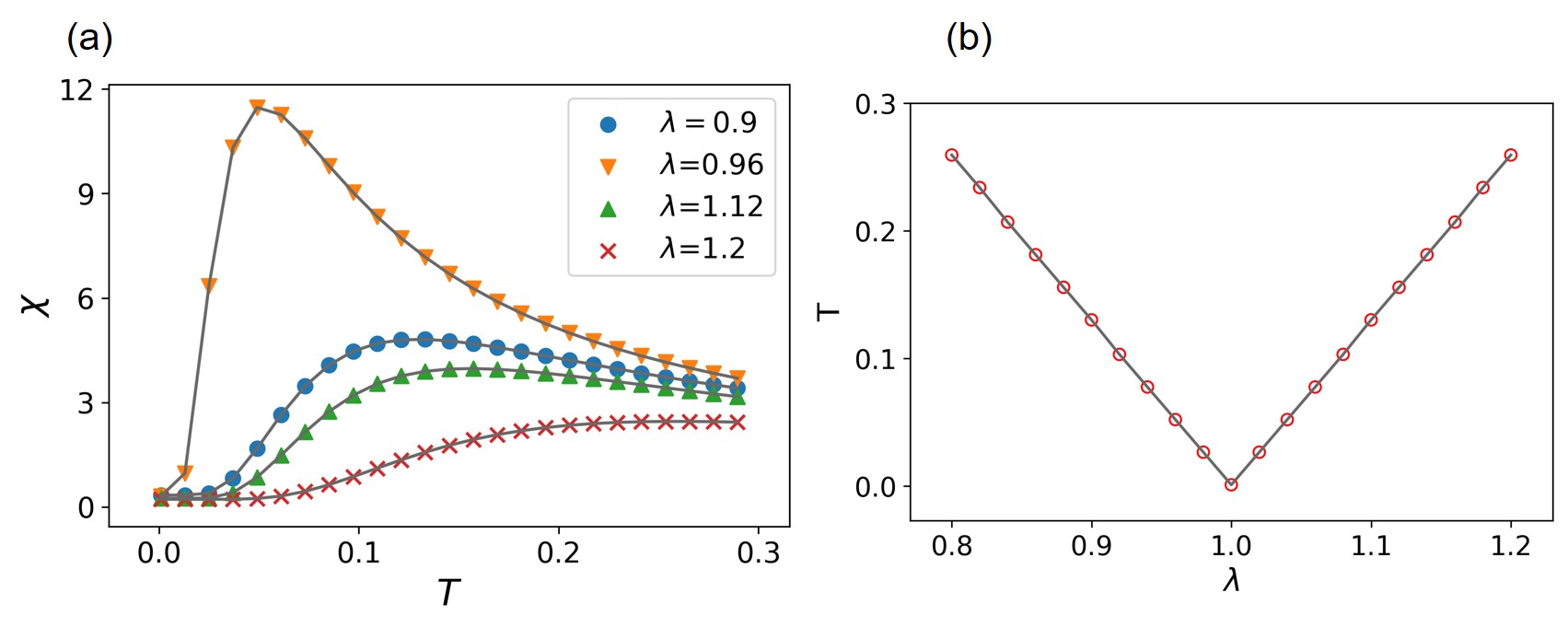}
	\caption{Locating the quantum critical regime by identifying the temperature crossover line. (a) For a given $\lambda$, the susceptibility $\chi$ is evaluated for varied temperatures~(only several $\lambda$ are shown); (b) The temperature crossover point for each $\lambda$ corresponds to a temperature that  $\xi$ is peaked in (a). For all scatters are simulation results and lines are exact results. } \label{Fig:regime_VQA}
\end{figure}

We then turn to the first goal of simulating the quantum critical regime: identifying the temperature crossover line. 
We chose the Kitaev model at $N=3$
as a minimal model to demonstrate the temperature crossover. The first step is to calculate the susceptibility $\chi(\beta,\lambda)$ according to the difference scheme in Eq.~\eqref{eq:diff_chi}, where $\delta\lambda=0.001$ is chosen. As shown in Fig.~\ref{Fig:regime_VQA}a, $\chi(\beta,\lambda)$ for a given $\lambda$ at different temperatures are calculated. The $\chi\sim T$ curve for each $\lambda$ is peaked at a temperature, which can be identified as a temperature crossover point. Based on Fig.~\ref{Fig:regime_VQA}a, the temperature crossover line can be obtained. As shown in Fig.~\ref{Fig:regime_VQA}b, the temperature crossover line obtained with VQA~(red dots) fits well with the exact one~(red line). 

\begin{figure}[htbp]
	\centering
	\includegraphics[width=1\linewidth]{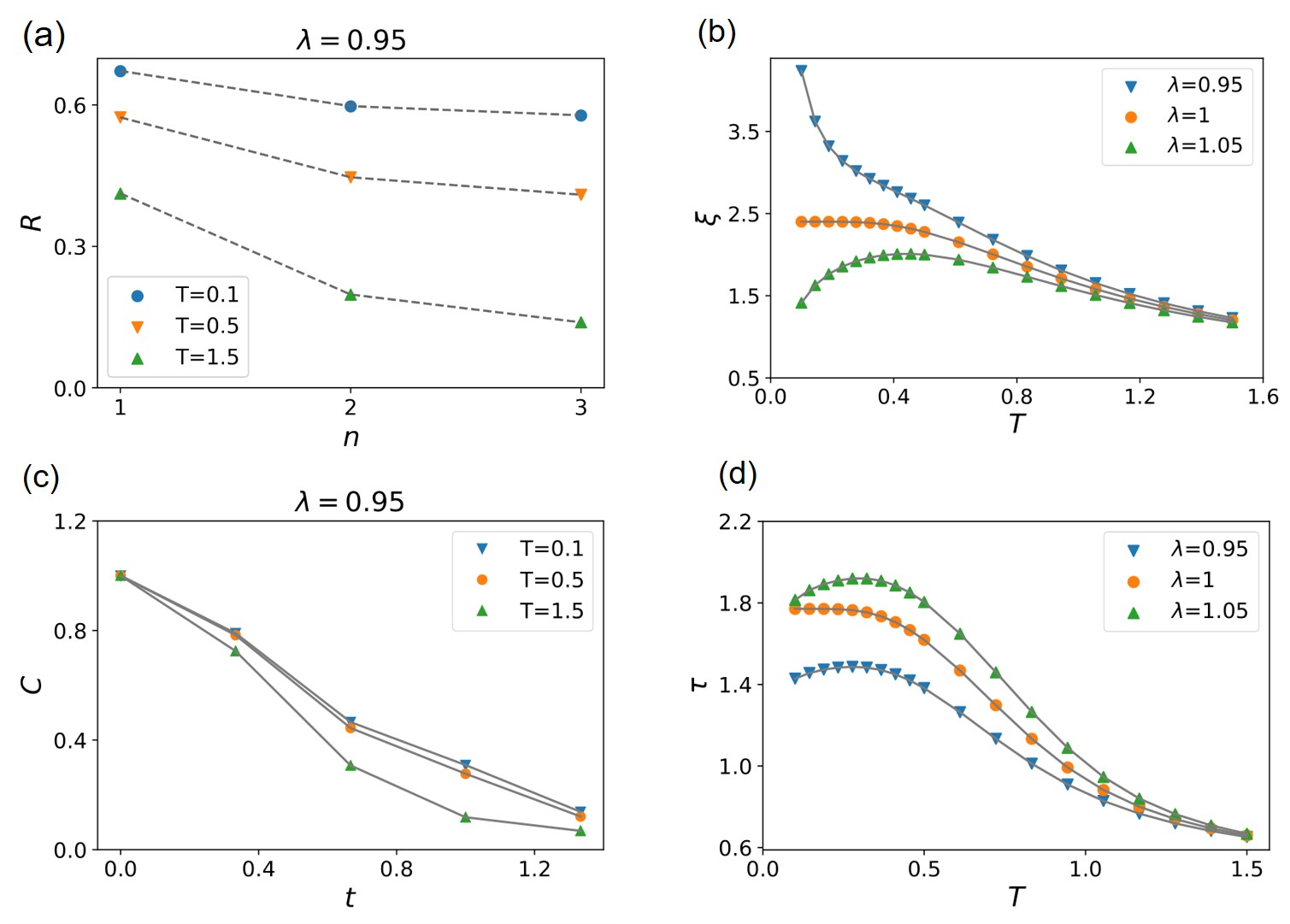}
	\caption{Dependence of the correlation length and the phase coherence time with temperature. (a) and (c) show the spatial correlation function $R$ with the spatial separation $n$ and the dynamical correlation function $C$ with the time period $t$, respectively. Only results of $T=0.1,0.5,1.5$ are shown; (b) shows the dependence of the correlation length $\chi$ with $T$ and (d) shows the dependence of the phase coherent time with $T$. For all,  simulation results marked by scatters are compared with exact results marked by lines.  }\label{Fig:correlation}
\end{figure}

The second goal is to investigate the scaling behavior in the quantum critical regime.  At this stage, the simulation is only limited to a very small size. We chose $N=6$. This allows $n$ in the spatial correlation function $R(n)$ can take $n=1,2,3$, which can be used for fitting the function $R(n)=ae^{-\frac{n}{\xi}}$ with three data. For a finite-size system, the largest time period $t$ in the dynamical correlation function $C(t)$ should be chosen so that $C(t)$ will not oscillate. As $t$ can be continuous, the number of data can be large to fit $C(t)=b e^{-t/\tau}$. With those in mind, $R(n)$ and $C(t)$ are measured for different $T$ and $\lambda$, which are shown in Fig.~\ref{Fig:correlation}a and Fig.~\ref{Fig:correlation}c respectively. Then, the correlation length and the phase coherent time are fitted with $R(n)$ and $C(t))$, respectively. In Fig.~\ref{Fig:correlation}b and Fig.~\ref{Fig:correlation}d, the relation $\xi\sim T$ and $\tau\sim T$ are presented. It can be seen that both are proportional to $T^{-1}$ in an intermediate regime of temperature. There are some mismatches at low temperatures. Remarkably at $\lambda=1$, it is expected that the correlation length should be divergent while the numeral simulation shows to be almost flat. This can be due to the finite-size effect. On the other hand, the dramatically increasing~(decreasing) of $\xi$ for $h=0.95$ is qualitatively consistent with theory. This corresponding to the semi-classical regimes, where $\chi$ should be exponentially growing with $T^{-1}$ for $\lambda<1$. Similarly, the phase coherent time ceases to increase with reducing temperature may be due to finite size. It is also observed a deviation of relations of $\xi\propto T^{-1}$ and $\tau\propto T^{-1}$ at high temperature. This can be explained that the high-temperature regime $T>J=1$ is governed by the lattice cutoff and thus no universal behavior can be expected. The above discussions suggest that the scaling behavior may be captured for an intermediate temperature regime. 

\section{Conclusions}\label{sec:conclusion}
In summary, we have proposed a variational quantum computing approach for simulating the quantum critical regime, using the Kitaev ring as a prototype model, by investigating the temperature crossover and the scaling behavior. 
The variational quantum algorithm adopts an ansatz that the free energy can be obtained free of the difficulty of measuring the entropy. By numeral simulation, we have shown that the variational quantum algorithm can identify the temperature crossover accurately. Moreover, we have shown that both the correlation length and the phase coherence time are proportional to the inverse temperature in an intermediate regime of temperature. Our work has paved the way for simulating finite-temperature critical systems on a quantum computer. 

\begin{acknowledgements}
This work was supported by the National Natural Science Foundation of China (Grant No.12005065) and the Guangdong Basic and Applied Basic Research Fund (Grant No.2021A1515010317).
\end{acknowledgements}

\bibliographystyle{apsrev4-2}
\bibliography{cRegime}

\end{document}